\documentclass[10pt,twocolumn]{IEEEtran}
\usepackage{graphicx}
\usepackage{amsmath}
\usepackage{amssymb}
\usepackage{amsbsy}
\usepackage{color}
\usepackage{amsmath}

\begin{document}


\title{\huge Decoupled Signal Detection for the Uplink of Large-Scale MIMO Systems in Heterogeneous Networks
}

\author{\IEEEauthorblockN{Leonel Ar\'evalo,
 Rodrigo C. de Lamare, Martin Haardt and Raimundo Sampaio-Neto}
\thanks{
Leonel Ar\'evalo, Rodrigo C. de Lamare and Raimundo Sampaio-Neto are with the Center for Studies in Telecommunications (CETUC), Pontifical Catholic University of Rio de Janeiro (PUC-Rio), Rio de Janeiro, Brazil (e-mail:\{leonel\_arevalo, delamare, raimundo\}@cetuc.puc-rio.br).}

\thanks{Martin Haardt is with the Communications Research Laboratory,  Ilmenau University of Technology, Ilmenau, Germany (e-mail: martin.haardt@tu-ilmenau.de).}

 \thanks{Parts of this paper have been published at CAMSAP 2015~\cite{ref84}.}

 \thanks{This work was supported in part by grants of CNPq and FAPERJ..}
 }
\maketitle

\begin{abstract}
Massive multiple-input multiple-output (MIMO) systems are strong candidates for future fifth generation (5G) heterogeneous cellular networks.
For 5G, a network densification with a high number of different classes of users and  data service requirements is expected. Such a large number of connected devices needs to be separated in order to allow the detection of the transmitted signals according to different data requirements.
In this paper, a decoupled signal detection (DSD) technique which allows the separation of  the uplink signals, for each user class, at the base station (BS)  is proposed for massive MIMO systems.
A mathematical signal model for massive MIMO systems with centralized and distributed antennas in heterogeneous networks is also developed.
The performance of the proposed DSD algorithm is evaluated and compared with existing detection schemes in a realistic scenario with distributed antennas. A sum-rate analysis and a computational cost study for DSD are also presented. Simulation results show an excellent performance of the proposed DSD algorithm when combined with linear and successive interference cancellation detection techniques.
\end{abstract}

\section{Introduction}
\label{intro}

\PARstart{L}{arge-scale} multiple-input multiple-output (MIMO)
systems, also known as massive MIMO, are a promising technology
which use a large number of antennas to serve a high number of user
terminals at the same time without requiring extra bandwidth
resources~\cite{ref5,ref81,ref53,ref54,zhang2015large,li2014distributed}.
This new greater scale version of traditional MIMO systems, where a
restricted number of antennas is used, is designed to exploit the
benefits of extra degrees of freedom obtained by the use of more
antennas~\cite{ref45}. Massive MIMO can increase the spectral
efficiency 10 times or more when compared with its
predecessor~\cite{ref57}. In the Long Term Evolution (LTE)
standard~\cite{ref46}, which uses traditional MIMO systems, with as
many as eight antennas at  the base station (BS), and operates in
the frequency-division duplex (FDD) mode, the users estimate the
channel response  and feed it back to the BS. For massive MIMO with
hundreds of antenna elements this might not be feasible, due to the
large number of channel coefficients that each user needs to
estimate being proportional to the number of the antennas at the BS.
In this paper we focus on the uplink, the reason is that the most
natural transmission mode to operate in massive MIMO is the
time-division duplexing (TDD) mode, where a reciprocity between the
uplink and downlink channels can be obtained, if we use appropriate
calibration techniques to combat the distortions induced by hardware
imperfections, since the base station can offer more processing
resources aimed at estimating the channel between users' terminals
and the BS. On the downlink it is possible to use different
precoding schemes to mitigate the interference for the received
signals at the mobile users. Such precoding schemes rely on the
channels estimated on the uplink.

One of the main research challenges of massive MIMO is to develop
computationally simple ways to process the large number of signals
received at the BS. The interference  between antennas and users,
propagation effects such as correlation, path loss and shadowing,
thermal noise and signal degradation due to the hardware
imperfections  need to be suppressed. Linear detection techniques
\cite{de2009adaptive,de2010reduced,de2010blind,de2011adaptive,cai2013adaptive,clarke2012transmit,peng2013adaptive,cai2015adaptive,peng2016adaptive}
such as maximum radio combining (MRC) and zero forcing (ZF) are a
good option in terms of computational complexity, however, their
performance is not compatible with the growing demand for high data
rates. The performance of linear detectors can be improved using
some nonlinear sub-optimal detector based on successive interference
cancellation
(SIC)~\cite{de2008minimum,ref2,gu2016joint,uchoa2016iterative},
e.g., multibranch SIC (MB-SIC)~\cite{ref14,ref7,de2013adaptive} and
multifeedback SIC
(MF-SIC)~\cite{ref22,cai2009adaptive,li2011multiple,li2012adaptive}.
However, SIC-based detectors have a considerable computational cost
for high-dimensional systems. If all signals from all active users
are coupled in the detection process, the BS could spend unnecessary
processing resources since some types of users might not require a
very high performance.

In the next generation of wireless communication
systems~\cite{ref47}, it is expected that a large number of users
with different configurations and requirements are connected to the
network. Therefore, it is necessary to design heterogeneous networks
capable of interconnecting the different user types with each
other~\cite{ref78}. The received signals from this large number of
connected devices such as metering equipment, sensors, environmental
monitoring devices, health care gadgets, security management
products, smart grid components, smart phones and tablets need to be
separated in order to detect the transmitted information according
to their different data requirements. In this context, distributed
antenna systems (DAS) with massive MIMO are a promising alternative
for the 5G cellular architecture~\cite{ref48}, where the BS will be
equipped with a large number of antennas and some remote antenna
arrays or radio heads will be distributed around the cell and
connected to the BS via optical fiber. The signals associated with
different remote antenna arrays are processed at the BS. DAS have
low path loss effects, improve the coverage and the spectral
efficiency~\cite{ref49}~\cite{ref62}. The energy consumption of
users is reduced and the transmission quality is improved due to the
shorter distances between users and some remote antenna arrays. For
this vision of 5G wireless networks, which includes a combination of
massive MIMO, heterogeneous networks, and distributed antenna
systems, efficient signal processing techniques at the BS are
necessary.

In this paper, we propose an algorithm for the uplink of massive MIMO systems to separate the combined received signal of all users at the BS into independent signals for each user class. The proposed decoupled signal detection (DSD) applies a decomposition into multiple independent single user class signals, where all users in a class have the same data requirements and a common complex modulation.
Assuming that the channel state information (CSI) was previously estimated,
DSD employs a common channel inversion and QR decomposition to decouple the received signal. Applying the proposed DSD algorithm, the computational cost of the  signal processing is reduced and it is possible to have flexibility on the detection procedures at the BS. A signal model for heterogeneous networks with different classes of users and an arbitrary configuration of centralized antenna systems (CAS) and distributed antenna systems (DAS) is also introduced in this paper.
A sum rate analysis and a computational complexity study for the proposed DSD are presented.
The performance of the proposed scheme is evaluated in a realistic scenario with propagation effects and compared with existing approaches.

The remainder of this paper is organized as follows.
In Section~\ref{sec:Preliminares} the proposed massive MIMO signal model is presented.
The proposed DSD scheme is presented in Section~III. The sum rate analysis for the DSD scheme is described in Section IV.
Section~V presents simulations results.
Section~VI gives the conclusions.

\section{Massive MIMO Signal Model}
\label{sec:Preliminares}

\begin{figure}
\begin{center}
\includegraphics[scale=0.17]{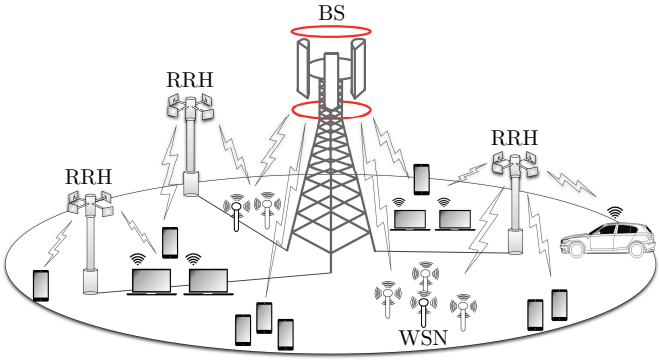}
\end{center}
\caption{Heterogeneous Wireless Network: base station (BS), remote radio head (RRH) and wireless sensor networks (WSN).}
\end{figure}

5G cellular networks will be able to support a large number and
diverse classes of devices, i.e., they will be heterogeneous, where
a single macro cell may need to support 10,000 or more low-rate
devices, along with its traditional high-rate mobile
users~\cite{ref86}.  This will require joint user classification
according to their data rate requirements. Thus, a user in a high
definition video call may have higher data rate requirements
compared to a mobile user transmitting voice. Then we can classify
them in different classes of users. In this section, a signal model
for heterogeneous networks with different classes of users and an
arbitrary configuration of CAS and DAS is presented.
We consider the uplink channel scenario of a massive MIMO system
with $N$ different classes of active users transmitting
simultaneously signals to one base station (BS) equipped with $D$
remote antenna arrays distributed around the cell and $N_B$ receive
antennas at the BS. Each remote array of antennas has $Q$ antennas
linked to the BS via wired links. Therefore, the total number of
receive antennas is $N_r=N_B+DQ$. The choice of $N_B$, $D$ and $Q$
is made based on the features of the network and type of application
scenario. For example, suppose that we have a city with a high
density of users or devices in the centre of a cell and sparsely
distributed users or devices in the remaining part of the cell. In
this case, we could use a number of centralized antennas to deal
with the high density of users and distributed antennas to serve the
remaining devices. The cardinality of the $n$-th user class $\mid
C_n\mid$, represents the number of users of the class $n$. The total
number of active users is given by $K=\sum_{n=1}^N\mid C_n\mid$. The
$k$-th user in the $n$-th user class transmits data divided into
$N_{t_{k,n}}$ sub-streams through $N_{t_{k,n}}$ antennas, where
$N_r\geqslant  N_t=\sum_{n=1}^N\sum_{k=1}^{\mid C_n\mid}N_{t_{k,n}}
$ and $N_t$ is the total number of transmit antennas. The received
signal vector at the BS from all active users in all user classes is
given by
\begin{equation}
\mathbf{y}=\sum_{n=1}^N\sum_{k=1}^{\mid C_n\mid}\mathbf{\Upsilon}_{k,n}\bar{\mathbf{H}}_{k,n}\mathbf{s}_{k,n}+\mathbf{n},\label{eq1}
\end{equation}
where $\mathbf{s}_{k,n}$ is the $N_{t_{k,n}}\times 1$ transmitted
signal vector, by the $k$-th user of the $n$-th user class, at one
time slot taken from a  complex constellation, denoted by
$\mathcal{A}=\{a_1, a_2, \ldots, a_O \}$. Each symbol has $M$ bits
and $O=2^M$. The vector $\mathbf{n}$ is an $N_r \times 1$ zero mean
complex circular symmetric Gaussian noise vector with covariance
matrix
$\mathbf{K}_\mathbf{n}=\mathbb{E}[\mathbf{n}\mathbf{n}^{\mathcal{H}}]=\sigma_n^2\mathbf{I}$.
Moreover, $\bar{\mathbf{H}}_{k,n}$ is the $N_r \times N_{t_{k,n}} $
channel matrix of the $k$-th user in the class $n$ with elements
$\bar{h}_{i,j}^{(k,n)}$ corresponding to the complex channel gain
from the $j$-th transmit antenna of the $k$-th user to the $i$-th
receive antenna. For the antenna elements located at the BS and at
each remote radio head, the $D+1$ sub-matrices of
$\bar{\mathbf{H}}_{k,n}=[(\bar{\mathbf{H}}_{k,n}^{(1)})^T,
(\bar{\mathbf{H}}_{k,n}^{(2)})^T, \ldots ,
(\bar{\mathbf{H}}_{k,n}^{(D+1)})^T]^T$ can be modeled using the
Kronecker channel model~\cite{ref11}, expressed by
\begin{equation}
\bar{\mathbf{H}}_{k,n}^{(j)}=(\mathbf{R}^{(j)}_{r})^{1/2}\mathbf{G}_{k,n}^{(j)} (\mathbf{R}_{t_{k,n}})^{1/2}
\end{equation}
where $\mathbf{G}_{k,n}^{(j)}$ has complex channel gains between the
$k$-th user and the $j$-th radio head, obtained from an independent
and identically distributed random fading model whose coefficients
are complex Gaussian random variables with zero mean and unit
variance. $\mathbf{R}_{r}^{(j)}$ and $\mathbf{R}_{t_{k,n}}$ denote
the receive correlation matrix of the $j$-th radio head and the
transmit correlation matrix, respectively. The components of the
correlation matrices $\mathbf{R}_{r}^{(j)}$ and
$\mathbf{R}_{t_{k,n}}$ are of the form:
\begin{equation}
\mathbf{R}=\begin{bmatrix}
1 & \rho & \rho^4 & \ldots & \rho^{(N_a-1)^2} \\ \rho & 1 & \rho & \ldots &
\vdots\\ \rho^4 & \rho & 1 & \vdots & \rho^4\\\vdots & \vdots & \vdots & \vdots & \vdots\\ \rho^{(N_a-1)^2} & \ldots & \rho^4 & \rho & 1
\end{bmatrix},
\end{equation}
where $N_a$ is the number of antennas and $\rho$ is the correlation
coefficient of neighboring antennas ($\rho=\rho_{t_x}$ for the transmit
antennas and $\rho=\rho_{r_x}$ for the receive antennas). Note that
$\rho =0$ represents an uncorrelated scenario and $\rho =1$ implies
a fully correlated scenario.
The $N_r\times N_r$ diagonal matrix $\mathbf{\Upsilon}_{k,n}$ represents the large-scale propagation effects for the user $k$ of the user class $n$, such as path loss and shadowing, given by
\begin{equation}
\mathbf{\Upsilon}_{k,n}=\text{diag}\Big{(}\underbrace{\gamma^1_{k,n}\ldots \gamma^1_{k,n}}_{N_B} \underbrace{\gamma^2_{k,n}\ldots \gamma^2_{k,n}}_{Q}\ldots  \underbrace{\gamma^{D+1}_{k,n}\ldots \gamma^{D+1}_{k,n}}_{Q} \Big{)},
\end{equation}
where the parameters $\gamma_{k,n}^j$ denote the large-scale propagation effects from the $k$-th user to the $j$-th radio head described by
\begin{equation}
\gamma_{k,n}^j=\alpha_{k,n}^j\beta_{k,n}^j, \,\,\,\,\,\,\, j=1,\ldots ,D+1.
\end{equation}
Here $\alpha_{k,n}^j$ is the distance based path-loss between each user and the radio heads which is calculated by
\begin{equation}
\alpha_{k,n}^j=\sqrt{\frac{L_{k,n}^j}{(d_{k,n}^j)^\tau}},
\end{equation}
where $L_{k,n}^j$ is the power path loss of the link associated with
the user and the $j$-th radio head, $d_{k,n}^j$ is the relative
distance between this user and the $j$-th radio head, $\tau$ is the
path loss exponent chosen between 2 and 4 depending on the
environment. The log normal random variable $\beta_{k,n}^j$ which
represents the shadowing between user $k$ and the receiver is given
by
\begin{equation}
\beta_{k,n}^j=10^{\frac{\mu_{k,s}\vartheta^j_{k,s}}{10}},
\end{equation}
where $\mu_{k,s}$ is the shadowing spread in dB and
$\vartheta^j_{k,s}$ corresponds to a real-valued Gaussian random
variable with zero mean and unit variance. Since the $N_r\times
N_{t_{k,n}}$ composite channel matrix includes large-scale and
small-scale fading effects it can be denoted as
$\mathbf{H}_{k,n}=\mathbf{\Upsilon}_{k,n}\bar{\mathbf{H}}_{k,n}$,
and the expression in (\ref{eq1}) can be written as
\begin{eqnarray}
\mathbf{y}&=& \sum_{n=1}^N \mathbf{H}_{n}\mathbf{s}_{n}+\mathbf{n},\label{eq3}
\end{eqnarray}
where $\mathbf{H}_n=[\mathbf{H}_{1,n} \,\,\, \mathbf{H}_{2,n} \ldots
\mathbf{H}_{\mid C_n\mid,n}]$ and $\mathbf{s}_n=[\mathbf{s}_{1,n}^T \,\,\,
\mathbf{s}_{2,n}^T \ldots \mathbf{s}_{\mid C_n\mid,n}^T ]^T$ represent the channel matrix and the transmitted symbol vector of all users in the class $n$, respectively. The received signal vector can be expressed more conveniently as
\begin{eqnarray}
\mathbf{y}&=&\mathbf{H}\mathbf{s}+\mathbf{n}, \label{eq2}
\end{eqnarray}
where $\mathbf{H}=[\mathbf{H}_1 \,\,\, \mathbf{H}_2 \ldots
\mathbf{H}_N ]$ and $\mathbf{s}=[\mathbf{s}_1^T \,\,\,
\mathbf{s}_2^T \ldots \mathbf{s}_N^T ]^T$. The symbol vector $\mathbf{s}$ of all $N$ user classes has zero mean and a covariance matrix
$\mathbf{K}_\mathbf{s}=\mathbb{E}[\mathbf{s}\mathbf{s}^\mathcal{H}]=\text{diag}(\mathbf{p})$, where the elements of the vector $\mathbf{p}$ are the signal power of each transmit antenna. To maintain a notational simplicity in the subsequent analysis, we assume that all antenna elements at the users transmit with the same average transmitted power $\sigma_s^2$, i.e., $\mathbf{K}_\mathbf{s}=\sigma_s^2\mathbf{I}$.
We assume that the channel matrix $\mathbf{H}$ was previously estimated at the BS. From~(\ref{eq2}) we can see that the signals arrive coupled at the BS. If we want to use different detection procedures for each user class according to its data requirements, we have to separate the received signal vector $\mathbf{y}$ into independent received signals for each user class.
For the system model presented in this work, when the number of remote radio heads is set to zero, i.e., $D=0$, the DAS architecture is reduced to the CAS scheme with $N_r=N_B$.

\section{Decoupled Signal Detection}
\label{sec:up-link_BD}

As presented in Section~\ref{sec:Preliminares}, in heterogeneous
networks different classes of users send parallel data streams,
through the massive MIMO channel operating with distributed
antennas, which arrive superposed at the BS. In this 5G context, we
need to separate the data streams for each category of users
efficiently. In this section, we describe the proposed decoupled
signal detection (DSD) which allows us to separate the received
signal of the $n$-th user class from the others. To this end, we
consider that the process of authentication, identification and
channel estimation was already made, i.e, the BS is able to identify
the users by classes according to their data requirements. Similar
approaches to the proposed DSD algorithm have been proposed for the
downlink, such as block diagonalization (BD) based
techniques~\cite{ref58}-\nocite{ ref59,
ref60},\cite{cai2015robust},\cite{Keke1},\cite{Keke2}\cite{ref61}
and other advanced precoding techniques \cite{Keke3}. However,
unlike prior work in which downlink BD is used, for the proposed DSD
scheme it is not necessary to use any precoding at the transmit
side. The receiver only needs to know the channels between users and
receive antennas. Moreover, the concept of separating the users with
respect to the classes in heterogeneous networks according to its
requirements is a new approach. The first steps to construct the
concept proposed in this paper were presented
in~\cite{ref84}\cite{ref85}.
The received signal vector~(\ref{eq3}) can be written as:
\begin{eqnarray}
\mathbf{y}&=& \mathbf{H}_{n}\mathbf{s}_{n}+\sum_{m=1, m\neq n}^N \mathbf{H}_{m}\mathbf{s}_{m}+\mathbf{n},\label{eq14}
\end{eqnarray}
where $\mathbf{H}_{n}$ and $\mathbf{s}_{n}$ are the channel matrix and the transmitted symbol vector for the $n$-th user class, respectively. From~(\ref{eq14}), we can see that the $n$-th user class has inter-user class interference.

\subsection{Proposed Decoupling Strategy}

To remove the presence of the other classes of users in the detection procedure for the $n$-th user class, we can employ a linear operation to project the received signal vector $\mathbf{y}$ onto the subspace orthogonal or almost orthogonal to the subspace generated by the signals of the interfering classes. In DSD, a matrix $\mathbf{A}_n$ is calculated employing a channel inversion method and a QR decomposition~\cite{ref82}~\cite{ref83}, in order to decouple the $n$-th user's class received signal from other user's class signals. To compute $\mathbf{A}_n$, we construct the matrix $\tilde{\mathbf{H}}_n$ excluding the channel matrix of the $n$-th user class in the following form:
\begin{equation}
\tilde{\mathbf{H}}_n=[\mathbf{H}_1 \ldots \mathbf{H}_{n-1}\,\,\, \mathbf{H}_{n+1} \ldots \mathbf{H}_N],
\label{eq5}
\end{equation}
where $\tilde{\mathbf{H}}_n \in \mathbb{C}^{N_r\times (N_t-N_{t_n})}$ and $N_{t_n}=\sum_{k=1}^{\mid C_n\mid}N_{t_{k,n}}$ is the number of transmit antennas in the $n$-th user class. After that, the objective is to obtain a matrix $\mathbf{A}_n$ that satisfies the following condition:
\begin{equation}
\mathbf{A}_n\tilde{\mathbf{H}}_n=\mathbf{0}, \,\,\,\,\, \forall n\in (1\ldots N).\label{eq7}
\end{equation}
To compute $\mathbf{A}_n$, DSD first computes the MMSE channel
inversion of the combined channel matrix $\mathbf{H}$ given by
\begin{eqnarray}
\mathbf{H}^\dagger &=&\mathbf{H}^\mathcal{H}(\mathbf{H}\mathbf{H}^\mathcal{H}+\sigma^2\mathbf{I})^{-1}\nonumber \\
&=& [(\ddot{\mathbf{H}}_1)^T \,\,\, (\ddot{\mathbf{H}}_2)^T \,\,\, \ldots \,\,\, (\ddot{\mathbf{H}}_N)^T]^T
\label{eq8}
\end{eqnarray}
where $\sigma^2=\sigma_n^2/\sigma_s^2$, $\mathbf{H}^\dagger \in \mathbb{C}^{N_t\times N_r}$ and $\ddot{\mathbf{H}}_n\in \mathbb{C}^{N_{t_n}\times N_r}$. Then, the matrix $\ddot{\mathbf{H}}_n$ is approximately in the null space of $\tilde{\mathbf{H}}_n$, that is
\begin{equation}
\ddot{\mathbf{H}}_n\tilde{\mathbf{H}}_n\approx \mathbf{0}, \,\,\,\,\, \forall n\in (1\ldots N).
\end{equation}
To decouple the $n$-th user group from the other user groups we
employ a QR decomposition as described by
\begin{equation}
\ddot{\mathbf{H}}_n=\mathbf{R}_n\mathbf{Q}_n, \,\,\,\,\, \forall n\in (1\ldots N),\label{eq9}
\end{equation}
where $\mathbf{R}_n \in \mathbb{C}^{N_{t_n}\times N_{t_n}}$ is an upper triangular matrix and $\mathbf{Q}_n \in \mathbb{C}^{N_{t_n}\times N_r}$ is a matrix with orthogonal rows and composed by approximately orthonormal basis vectors of the left null space of $\tilde{\mathbf{H}}_n$. Then we have
\begin{equation}
\mathbf{Q}_n\tilde{\mathbf{H}}_n\approx \mathbf{0}.\label{eq6}
\end{equation}
From~(\ref{eq6}), we can see that $\mathbf{Q}_n$ is a good approximation for $\mathbf{A}_n$ in~(\ref{eq7}). Using $\mathbf{A}_n=\mathbf{Q}_n$ as a linear combination with the received signal vector in~(\ref{eq14}), we have
\begin{eqnarray}
\mathbf{y}_n&=&\mathbf{A}_n\mathbf{y} \nonumber \\
&=&\mathbf{Q}_n\mathbf{H}_{n}\mathbf{s}_{n}+\mathbf{Q}_n\sum_{\substack{ m=1\\ m\neq n}}^N \mathbf{H}_{m}\mathbf{s}_{m}+\mathbf{Q}_n\mathbf{n},\label{eq13}
\end{eqnarray}
where $\mathbf{y}_n\in \mathbb{C}^{N_{t_n}\times 1}$ is the equivalent received signal vector for the user class $n$ and the term $\mathbf{Q}_n{\fontsize{2.5}{4}\sum_{\substack{ m=1\\ m\neq n}}^{N}} \mathbf{H}_{m}\mathbf{s}_{m}\approx \mathbf{0}$ represents the residual inter-user class interference. Then, we can transform the received signal vector into parallel single-user class signals as described by
\begin{equation}
\mathbf{y}_n=\check{\mathbf{H}}_n\mathbf{s}_n+\mathbf{n}_n, \,\,\,\,\, \forall n\in (1\ldots N),\label{eq12}
\end{equation}
where $\check{\mathbf{H}}_n=\mathbf{Q}_n\mathbf{H}_{n}\in
\mathbb{C}^{N_{t_n}\times N_{t_n}}$ is the equivalent channel matrix
of the $n$-th user class after DSD and
$\mathbf{n}_n=\mathbf{Q}_n{\fontsize{2.5}{4}\sum_{\substack{ m=1\\
m\neq n}}^{N}}
\mathbf{H}_{m}\mathbf{s}_{m}+\mathbf{Q}_n\mathbf{n}\in
\mathbb{C}^{N_{t_n}\times 1}$ is the equivalent noise vector.

Another option to compute a basis for the left null space of
$\tilde{\mathbf{H}}_n$ is performing the SVD transformation
$\tilde{\mathbf{H}}_n=\tilde{\mathbf{U}}_n \tilde{\mathbf{\Sigma}}_n
\tilde{\mathbf{V}}_n^\mathcal{H}$, where $\tilde{\mathbf{\Sigma}}_n
\in \mathbb{C}^{N_r \times (N_t-N_{t_n})}$  is a rectangular
diagonal matrix with the singular values of $\tilde{\mathbf{H}}_n$
on the diagonal, $\tilde{\mathbf{U}}_n \in \mathbb{C}^{N_r \times
N_r}$ and $ \tilde{\mathbf{V}}_n^\mathcal{H} \in
\mathbb{C}^{(N_t-N_{t_n}) \times (N_t-N_{t_n})}$ are unitary
matrices. If $r_n$ is the rank of $\tilde{\mathbf{H}}_n$, that
corresponds to the number of non-zero singular values, i.e.,
$r_n={\rm rank}(\tilde{\mathbf{H}}_n)\leq  N_t-N_{t_n} $, the SVD
can be expressed equivalently as:
\begin{equation}
\tilde{\mathbf{H}}_n=[\tilde{\mathbf{U}}_{1, n} \,\,\, \tilde{\mathbf{U}}_{0, n}]\,\,\, \tilde{\mathbf{\Sigma}}_n\,\,\, [\tilde{\mathbf{V}}_{1, n}\,\,\, \tilde{\mathbf{V}}_{0, n}]^\mathcal{H},\label{eq10}
\end{equation}

where~$\tilde{\mathbf{U}}_{0, n}~\in~\mathbb{C}^{N_r \times
(N_r-r_n)}$~and~$\tilde{\mathbf{V}}^\mathcal{H}_{0,
n}~\in~\mathbb{C}^{(N_t-N_{t_n}-r_n)\times (N_t-N_{t_n})}$ form  an
orthogonal basis for the left null space and the null space of
$\tilde{\mathbf{H}}_n$, respectively. Then, the alternative solution
for~(\ref{eq7}) could be:
\begin{equation}
\mathbf{A}_n=\tilde{\mathbf{U}}_{0, n}^\mathcal{H}.
\end{equation}

Although the matrix $\tilde{\mathbf{U}}_{0, n}^\mathcal{H}$
eliminates the inter-user class interference effectively, i.e.,
$\tilde{\mathbf{U}}_{0,
n}^\mathcal{H}{\fontsize{2.5}{4}\sum_{\substack{ m=1\\ m\neq
n}}^{N}} \mathbf{H}_{m}\mathbf{s}_{m}=\mathbf{0}$, the linear
combination $\mathbf{y}_n=\tilde{\mathbf{U}}_{0,
n}^\mathcal{H}\mathbf{y}$ presents a noise enhancement  in the
detection procedures which reduces the performance. In the first
approach, when we use $\mathbf{y}_n=\mathbf{Q}_n\mathbf{y}$, the
noise enhancement in the detection procedures is mitigated due to
the fact that the equivalent noise vector has a lower dimension when
compared with the other option. Thus, the noise enhancement is
reduced which improves the performance, even in the presence of
residual interference. In addition, the equivalent channel matrix
$\check{\mathbf{H}}_n^1=\tilde{\mathbf{U}}_{0,
n}^\mathcal{H}\mathbf{H}_n$ has dimensions $(N_r-r)\times N_t$
against the matrix $\check{\mathbf{H}}_n^2=\mathbf{Q}_n\mathbf{H}_n$
which has dimensions $N_{t_n}\times N_{t_n}$. For this reason the
computational complexity of the detector is lower if we use the
matrix $\mathbf{Q}_n$ to decouple the received signal vector.

The fact that we obtain a square equivalent channel matrix also
allows the possibility of using lattice reduction (LR) based
detectors which have a better performance for square channel
matrices~\cite{ref55}. Further, the computational complexity to
compute the channel inversion~(\ref{eq8}) and $N$ QR
decompositions~(\ref{eq9}) of matrices with dimensions
$N_{t_n}\times N_r$ is much lower than the computational cost of
computing $N$ SVD transformations~(\ref{eq10}) of matrices with
dimensions $N_r\times (N_t-N_{t_n})$. For these reasons, in this
paper we will focus on the first alternative.

As it will be presented in next section, the equivalent received
signal vector in~(\ref{eq12}) shows that the process
in~(\ref{eq5})-(\ref{eq13}) is an effective algorithm to separate
the user classes at the BS and we can consider the data stream of
the $n$-th user class as independent of the received signals of the
other user classes. In practice, this allows the possibility of
using different transmission and reception schemes for each user
class. We can now implement the traditional detectors for each class
of users separately  which also allows the possibility of using more
complex detection schemes due to the reduction of the dimensions of
the matrices that need to be processed. The description of the
proposed DSD algorithm is presented in Algorithm~1.

\subsection{Detection Algorithms}

In this subsection, we examine signal detection algorithms for
massive MIMO in heterogeneous networks. To detect the data stream
for  each class of users independently, we assume that the DSD
algorithm described in~Algorithm~1 was previously employed.


\subsubsection{Linear Detectors}

In linear detectors, the equivalent received signal vector for the
$n$-th user class $\mathbf{y}_{n} \in \mathbb{C}^{N_{t_n}\times 1}$
is processed by a linear filter to eliminate the channel
effects~\cite{ref26}. The two linear detectors considered here are
given by
\begin{equation}
\mathbf{W}_n^\chi=
\begin{cases}
    (\check{\mathbf{H}}_n^\mathcal{H}\check{\mathbf{H}}_n)^{-1}\check{\mathbf{H}}_n^\mathcal{H},& \chi= \text{Zero Forcing }\\
    (\check{\mathbf{H}}_n^\mathcal{H}\check{\mathbf{H}}_n+\sigma^2\mathbf{I})^{-1}\check{\mathbf{H}}_n^\mathcal{H},              & \chi=\text{MMSE}
\end{cases}\label{eq20}
\end{equation}
where $\sigma^2=\sigma_n^2/\sigma_s^2$ and $\check{\mathbf{H}}_n\in
\mathbb{C}^{ N_{t_n}\times N_{t_n}}$ is the equivalent channel
matrix of the $n$-th user class. Note that for the MMSE detector, we
consider the autocorrelation matrix of the equivalent noise vector
as $\mathbf{K}_{\mathbf{n}_n}\approx \sigma_n^2\mathbf{I}$. As the
residual interference is very small, an excellent performance can be
obtained  with this approximation. The linear hard decision of
$\mathbf{s}_n$ is carried out as follows:
\begin{equation}
\hat{\mathbf{s}}_n=\mathbb{C}(\mathbf{W}_n^\chi\mathbf{y}_{n}),
\end{equation}
where the function $\mathbb{C}(x)$ returns the point of the complex
signal constellation closest to $x$. The linear detectors have a
lower computational complexity when compared with the non-linear
detectors. However, due to the impact of interference and noise,
linear detectors offer a limited performance.

\subsubsection{Successive Interference Cancellation (SIC)}

The SIC detector for the $n$-th user class in~(\ref{eq12}) consists
of a bank of linear detectors, each detects a selected component
$s_{n,i}$ of $\mathbf{s}_n$. The component obtained by the first
detector is used to reconstruct the corresponding signal vector
which is then subtracted from the equivalent received signal to
further reduce the interference in the input to the next linear
receive filter. The successively canceled received data vector that
follows a chosen ordering in the $i$-th stage is given by
\begin{equation}
\mathbf{y}_{n,i}=\mathbf{y}_{n}-\sum_{j=1}^{i-1}\check{\mathbf{h}}_{n,j}\hat{s}_{n,j},
\end{equation}
where $\check{\mathbf{h}}_{n,j}$ correspond to the columns of the channel matrix $\check{\mathbf{H}}_n$ and $\hat{s}_{n,j}$ is the estimated symbol obtained at the output of the $j$-th linear detector.
\subsubsection{Multiple-Branch SIC Detection}
In the multi-branch scheme~\cite{ref7} for the $n$-th user class, different orderings are explored for SIC, each ordering is referred to as a branch, so that a detector with
$L$ branches produces a set of $L$ estimated vectors. Each branch
uses a column permutation matrix $\mathbf{P}_n$.  The estimate of the
signal vector of branch $l$, $\mathbf{\hat{x}}_n^{(l)}$, is obtained using
a SIC receiver based on a new channel matrix $\check{\mathbf{H}}_n^{(l)}
=\check{\mathbf{H}}_n\mathbf{P}_n^{(l)}$. The order of the estimated symbols is
rearranged to the original order by
\begin{equation}
\mathbf{\hat{s}}_n^{(l)}=\mathbf{P}_n^{(l)}\mathbf{\hat{x}}_n^{(l)},  \,\,\,\,\,\,\, l=1, \ldots, L.\label{eq4}
\end{equation}
A higher detection diversity can be obtained by selecting the most
likely symbol vector based on the ML selection rule, that is
\begin{equation}
\mathbf{\hat{s}}_n= \text{arg min}\parallel \mathbf{y}_{n}-\check{\mathbf{H}}_n\mathbf{\hat{s}}_n^l \parallel^2, \hspace{0.5cm} l=1,\ldots,L.
\end{equation}

Other detectors could be used with the proposed DSD technique and this is up to the designer to choose the detector.
\section{Sum-Rate Analysis}
\label{Sec: Sum Rate}
\newcounter{mytempeqncnt}
\begin{figure*}[!t]
\normalsize
\setcounter{mytempeqncnt}{\value{equation}}
\setcounter{equation}{34}
\begin{equation}
\label{eq22}
R_{k,n}=\mathbb{E}\Bigg{[}\log_2 \Bigg{(}1+\frac{\sigma_s^2\mid \mathbf{w}_{k,n}\check{\mathbf{h}}_{k,n}\mid^2}{\sigma_s^2\sum_{\substack{ j=1\\ j\neq k}}^{\mid C_n\mid}\mid\mathbf{w}_{k,n}\check{\mathbf{h}}_{j,n}\mid^2+\mid\mathbf{w}_{k,n}[\sigma_s^2 \mathbf{Q}_n(\sum_{\substack{ m=1\\ m\neq n}}^{N} \mathbf{H}_{m}\mathbf{H}_{m}^\mathcal{H})\mathbf{Q}_n^\mathcal{H}+\sigma_n^2\mathbf{I}]\mathbf{w}_{k,n}^\mathcal{H}\mid}\Bigg{)}\Bigg{]} ,
\end{equation}
\setcounter{equation}{\value{mytempeqncnt}}
\hrulefill
\vspace*{4pt}
\end{figure*}
In this section, a performance analysis for the proposed DSD scheme is presented in terms of the sum rate. We consider that the channel matrix $\mathbf{H}$ was previously estimated at the BS, assume Gaussian signalling and that the received signal vector was decoupled for each user class. Considering the received signal vector as presented in~(\ref{eq12}), the sum rate~\cite{ref79} that DSD can offer is defined as
\begin{equation}
R=\sum_{n=1}^{N}\log_2 \frac{\det(\mathbf{K}_{\mathbf{y}_n})}{\det(\mathbf{K}_{\mathbf{n}_n})},\label{eq15}
\end{equation}
where $\mathbf{K}_{\mathbf{y}_n}$ and $\mathbf{K}_{\mathbf{n}_n}$ are the autocorrelation matrix of the equivalent received signal vector and the equivalent noise vector of the $n$-th user class, respectively. It is easy to show that~(\ref{eq15}), can be expressed as
\begin{equation}
R=\sum_{n=1}^N \log_2 \det(\mathbf{I}+\mathbf{B}_n\mathbf{B}_n^\mathcal{H}),
\end{equation}
where $\mathbf{B}_n=\mathbf{K}_{\mathbf{n}_n}^{-1/2}\check{\mathbf{H}}_s\mathbf{K}_{\mathbf{s}_n}^{1/2} \in \mathbb{C}^{N_{t_n}\times N_{t_n}}$.
As $\mathbf{B}_n\mathbf{B}_n^\mathcal{H}$ is a Hermitian symmetric positive-definite square matrix, we have
\begin{equation}
\mathbf{B}_n\mathbf{B}_n^\mathcal{H}=\bar{\mathbf{Q}}_n\mathbf{\Lambda}_n\bar{\mathbf{Q}}_n^\mathcal{H},
\end{equation}
where $\bar{\mathbf{Q}}_n$ is a square unitary matrix, $\bar{\mathbf{Q}}_n\bar{\mathbf{Q}}_n^\mathcal{H}=\mathbf{I}$, and $\mathbf{\Lambda}_n$ is a diagonal matrix whose diagonal elements are the eigenvalues of the matrix $\mathbf{B}_n\mathbf{B}_n^\mathcal{H}$.
Then, the reliable sum rate that the system can offer is
\begin{equation}
R=\sum_{n=1}^N\sum_{i=1}^{N_{t_n}}\log_2 (1+\lambda_{i,n}).\label{eq16}
\end{equation}
Note that the eigenvalues $\lambda_{i,n}$ in~(\ref{eq16}) can be obtained computing the eigenvalues of  $\mathbf{B}_n^\mathcal{H}\mathbf{B}_n$.
As mentioned before, for notational simplicity we assume that $\mathbf{K}_{\mathbf{s}_n}=\sigma_s^2\mathbf{I}$. When the DSD algorithm is applied, the equivalent noise vector for the $n$-th user class $\mathbf{n}_n={\fontsize{2.5}{4}\mathbf{Q}_n\sum_{\substack{ m=1\\ m\neq n}}^{N} \mathbf{H}_{m}\mathbf{s}_{m}+\mathbf{Q}_n\mathbf{n}}\in \mathbb{C}^{N_{t_n}\times 1}$ is not white due to the residual inter-user class interference. Then its autocorrelation matrix is given by
\begin{equation}
\mathbf{K}_{\mathbf{n}_n}=\mathbb{E}[\mathbf{n}_n\mathbf{n}_n^\mathcal{H}]=\sigma_s^2 \mathbf{Q}_n\Big(\sum_{\substack{ m=1\\ m\neq n}}^{N} \mathbf{H}_{m}\mathbf{H}_{m}^\mathcal{H}\Big)\mathbf{Q}_n^\mathcal{H}+\sigma_n^2\mathbf{I}.\label{eq23}
\end{equation}
Finally, the eigenvalues $\lambda_{i,n}$ in~(\ref{eq16}) are obtained from the eigenvalues of matrix $\mathbf{B}_n^\mathcal{H}\mathbf{B}_n$ given by
\begin{equation}
\mathbf{B}_n^\mathcal{H}\mathbf{B}_n=\frac{\sigma_s^2}{\sigma_n^2}\check{\mathbf{H}}_n^\mathcal{H} \Big[\frac{\sigma_s^2}{\sigma_n^2} \mathbf{Q}_n\Big(\sum_{\substack{ m=1\\ m\neq n}}^{N} \mathbf{H}_{m}\mathbf{H}_{m}^\mathcal{H}\Big)\mathbf{Q}_n^\mathcal{H}+\mathbf{I}\Big]^{-1}\check{\mathbf{H}}_n.\label{eq18}
\end{equation}

To compute the sum rate for the received signal vector in~(\ref{eq2}) when the detection is perform for all user classes together, we suppress the index $n$ from the above analysis and considering that  $\mathbf{K}_\mathbf{s}=\sigma^2_s\mathbf{I}$ and $\mathbf{K}_\mathbf{n}=\sigma_n^2\mathbf{I}$, we get the well-known expression:
\begin{equation}
\bar{R}=\sum_{i=1}^{N_t}\log_2(1+\lambda_i),\label{eq17}
\end{equation}
where the values $\lambda_i$ are the eigenvalues of the matrix $\mathbf{B}^\mathcal{H}\mathbf{B}=\frac{\sigma^2_s}{\sigma_n^2}\mathbf{H}^\mathcal{H}\mathbf{H}$~\cite{ref79}. In Appendix~A, we show that, as well as the sum rate in~(\ref{eq17}) when all user classes are detected together, the sum rate in~(\ref{eq16}) for the proposed DSD algorithm is independent of the detection procedure. However, the lower bound on the achievable uplink sum rate obtained by using linear detectors is different for each detector~\cite{ref80}. In order to analyze the behavior of the lower bound on the sum rate for the proposed DSD scheme we consider that a linear detector according to~(\ref{eq20}) is applied to the equivalent received signal vector~(\ref{eq12}) to detect the transmitted symbol vector of the user class $n$, then we have
\begin{equation}
\tilde{\mathbf{y}}_n=\mathbf{W}_n \mathbf{y}_n=\mathbf{W}_n \check{\mathbf{H}}_n\mathbf{s}_n+\mathbf{W}_n \mathbf{n}_n.\label{24}
\end{equation}
Taking the $k$-th element of $\tilde{\mathbf{y}}_n$ we have
\begin{equation}
\tilde{y}_{k,n}=\mathbf{w}_{k,n}\check{\mathbf{h}}_{k,n}s_{k,n}+\sum_{\substack{ j=1\\ j\neq k}}^{\mid C_n\mid}\mathbf{w}_{k,n}\check{\mathbf{h}}_{j,n}s_{j,n}+\mathbf{w}_{k,n}\mathbf{n}_n,\label{eq21}
\end{equation}
where $\mathbf{w}_{k,n}$ is the $k$-th row of $\mathbf{W}_n$. Modeling the noise interference, the inter-user class interference and the inter-user interference in the user class $n$ in~(\ref{eq21}) as additive Gaussian noise independent of $s_{k,n}$, considering~(\ref{eq23})  and that the channel is ergodic so that each codeword spans over a large number of realizations, we obtain the lower bound on the achievable rate for the DSD algorithm with linear detectors as described in~(\ref{eq22}).
Then, the lower bound on the achievable rate for the entire system is given by
\newcounter{Mytempeqncnt}
\setcounter{Mytempeqncnt}{\value{equation}}
\setcounter{equation}{35}
\begin{equation}
R=\sum_{n=1}^N\sum_{k=1}^{\mid C_n\mid}R_{k,n}.
\end{equation}

For the SIC receiver, each stream is filtered by a linear detector and then, its contribution is subtracted from the received signal to improve the subsequent detection. For each layer the linear filter is recalculated. The performance of SIC detectors can be improved if we choose the cancellation order as a function of the SINR at the output of the linear detector in each layer. The lower bound for the sum rate of the proposed DSD algorithm when a SIC detector is used for each user class, could be calculated updating the expression~(\ref{eq22}) in each layer, i.e., the values of $\mathbf{w}_{k,n}$ are recalculated for each detected stream.

\section{Numerical Results}
\label{sec:results}
In this section, we evaluate the performance of the proposed DSD algorithm with different detectors in terms of the sum rate and the BER via simulations. Moreover, the computational complexity of the proposed and existing algorithms is also evaluated.

\subsection{Sum Rate}
To evaluate the analytic results obtained in Section~\ref{Sec: Sum Rate}, the sum rate and the lower bounds for the proposed DSD algorithm with different detection schemes will be evaluated considering CAS and DAS configuration assuming perfect CSI.
For the CAS configuration, we employ $L_k=0.7$, $\tau=2$, the distance $d_k$ to the BS is obtained from a uniform discrete random variable distributed between 0.1 and 0.99, the shadowing spread is $\sigma_k=3$ dB and the transmit and receive correlation coefficients are $\rho_{rx}=0.2$ and $\rho_{tx}=0.4$ (when $N_{t_{k,n}}>1$), respectively.
For DAS configurations, we consider a densely populated cell, where a fraction of the active users are in the centre of the cell and the remaining users are in other locations of the cell. We explore different particular values for the fraction of users in the centre  and around the cell. Based on that, we choose specific values for $N_B$, $D$ and $Q$.
For the DAS configuration, we also consider $L_{k,j}$ taken from a uniform random variable distributed between 0.7 and 1, $\tau=2$, the distance $d_{k,j}$ for each link to an antenna is obtained from a uniform discrete random variable  distributed between 0.1 and 0.5, the shadowing spread is $\sigma_{k,j}=3$ dB and the transmit and receive correlation coefficients are $\rho_{rx}=0.2$ and $\rho_{tx}=0.4$ , respectively.

\begin{figure}[h]
\begin{center}
\includegraphics[scale=0.62]{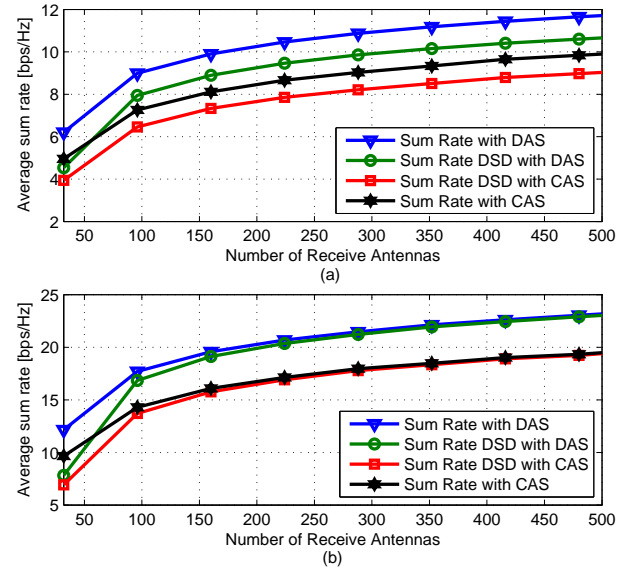}
\end{center}
\caption{(a) Sum Rate vs Number of Receive antennas. N=4, $\mid C_n\mid =8$ users per class, $N_{t_{k,n}}=1$ antennas per user.
(b) Sum Rate vs Number of Receive antennas. N=16, $\mid C_n\mid =1$ users per class, $N_{t_{k,n}}=2$ antennas per user.}
\label{R2}
\end{figure}

\begin{figure}[h]
\begin{center}
\includegraphics[scale=0.62]{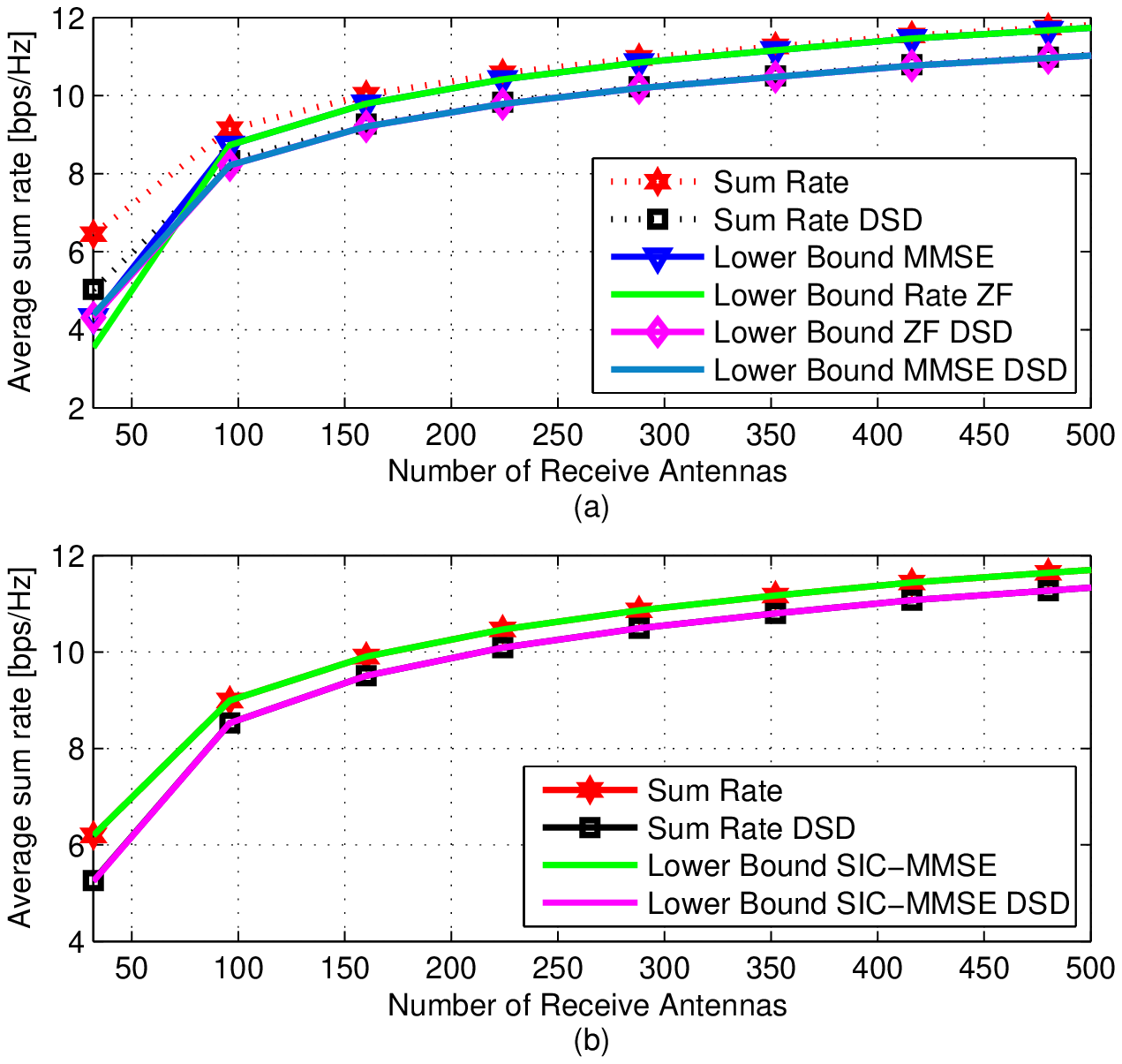}
\end{center}
\caption{(a) Sum Rate vs Number of Receive antennas. N=3, $\mid C_n\mid =10$ users per class, $N_{t_{k,n}}=1$ antennas per user.
(b) Sum Rate vs Number of Receive antennas. N=2, $\mid C_n\mid =16$ users per class, $N_{t_{k,n}}=1$ antennas per user.}
\label{R3}
\end{figure}

\begin{figure}[h]
\begin{center}
\includegraphics[scale=0.62]{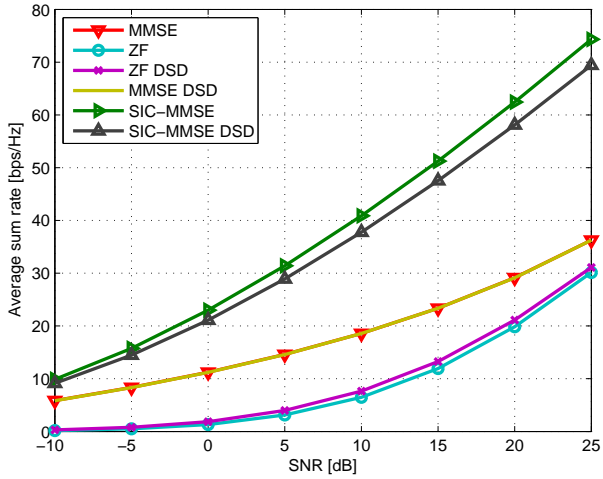}
\end{center}
\caption{Sum Rate vs SNR. N=8, $\mid C_n\mid =1$ users per class, $N_{t_{k,n}}=8$ antennas per user. We consider DAS configuration with $N_B=96$, $D=4$ and $Q=8$.}
\label{R4}
\end{figure}

In Fig.~\ref{R2} we evaluate the sum rate in two different scenarios for the users requirements. In both cases, we fix the SNR=10 dB and increase the number of receive antennas. For the DAS configuration, we consider $N_B=1/2N_r$ antennas at the BS. We also consider $D=4$ arrays of antennas distributed around the cell, each equipped with $Q=1/8N_r$ antennas.
For Fig.~\ref{R2}~(a) we consider $N=4$ classes of users with $\mid C_n \mid=8$ users each and $N_{t_{k,n}}=1$ antennas per user. We can see that the sum rate of the proposed DSD algorithm is close to the sum rate of the traditional MIMO system and with a low computational complexity on the detection procedures as will be shown in the next subsection. For Fig.~\ref{R2}~(b), we consider 16 active users in the system and that we need to detect each user independently, i.e.,  $N=16$ classes of users with $\mid C_n \mid=1$ users at each class and $N_{t_{k,n}}=2$ antennas per user. Under these conditions, the sum rate of the proposed scheme reaches the sum rate of the traditional MIMO system, especially for a large number of receive antennas. As expected, from the plot in Fig.~\ref{R2}, we can see that the sum rate for DAS is higher than that of the CAS configuration.

In Fig.~\ref{R3}, we compare the lower bound on the sum rate for different detectors such as, ZF, MMSE and SIC. We consider the DAS configuration under the same conditions as in the previous experiment. For Fig.~\ref{R3}~(a) we consider $N=3$ classes of users with $\mid C_n \mid=10$ users at each class and $N_{t_{k,n}}=1$ antennas per user. We can see from the plot, that similarly to the traditional MIMO systems, the lower bound on the sum rate for ZF and MMSE achieves the sum rate when $N_r$ grows. For Fig.~\ref{R3}~(b) we consider $N=2$ classes of users with $\mid C_n \mid=16$ users at each class and $N_{t_{k,n}}=1$ antennas per user. We can see that the SIC-MMSE achieves the sum rate and it could be considered optimal in terms of sum rate.

In Fig.~\ref{R4}, we compare the lower bound sum rate versus SNR. We consider $N=8$ classes of users with $\mid C_n \mid=1$ user in each class and $N_{t_{k,n}}=8$ antennas per user transmitting with high correlation between antennas $\rho_{tx}=0.85$. We consider the DAS configuration with $N_B=96$, $D=4$ and $Q=8$. We can see from the plot that the lower bounds for the proposed DSD algorithms are very close to the lower bounds when the detection procedure is carried out together for all users.


\subsection{Computational Complexity Analysis}
In this subsection, the computational complexity of the proposed DSD algorithm is evaluated and compared with the traditional coupled detection schemes, when all user classes are detected together, by counting the number of floating point operations (FLOPs) per received vector $\mathbf{y}$. Different detection schemes are considered such as MMSE, SIC and MB-SIC. The SIC based receivers all use MMSE detection. Furthermore, the single-branch SIC and the first branch of the MB-SIC employ norm-based ordering.
We consider QPSK modulation, however the computational cost in these detectors does not change significantly with the modulation order.
The number of FLOPs for the complex QR decomposition of an $N_{t_n}\times N_r$ matrix is given in~\cite{ref77} as $16(N_r^2N_{t_n}-N_{t_n}^2N_r+1/3N_{t_n}^3)$.
To compute the number of FLOPs required for the remaining operations, we use the Light-speed Matlab toolbox~\cite{ref15}.

\begin{figure}[]
\begin{center}
\includegraphics[scale=0.6265]{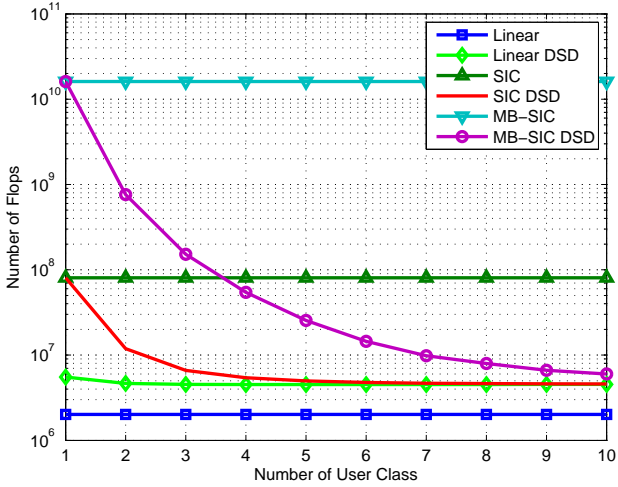}
\end{center}
\caption{Computational complexity versus number of user class, $K=100$ active users, $N_{t_{k,n}}=2$ antennas per user, $N_r=200$ receive antennas.}
\label{CC1}
\end{figure}

\begin{figure}[]
\begin{center}
\includegraphics[scale=0.62]{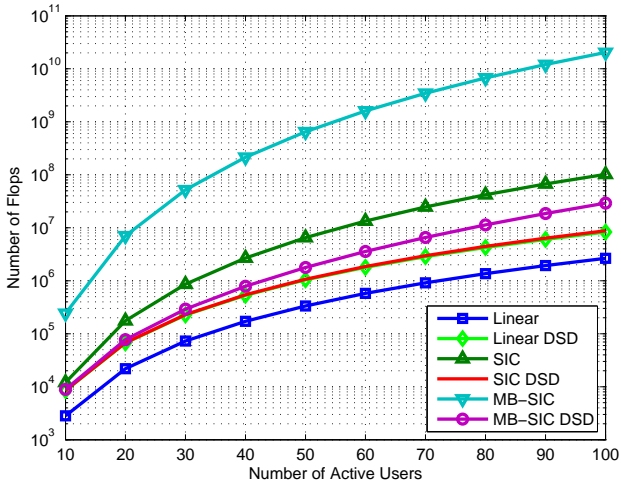}
\end{center}
\caption{Computational complexity versus number of active user, $N=5$ classes of user, $N_{t_{k,n}}=2$ antennas per user, $N_t=KN_{t_{k,n}}$ transmit antennas and $N_r=3N_t$ receive antennas.}
\label{CC2}
\end{figure}

\begin{figure}[]
\begin{center}
\includegraphics[scale=0.618]{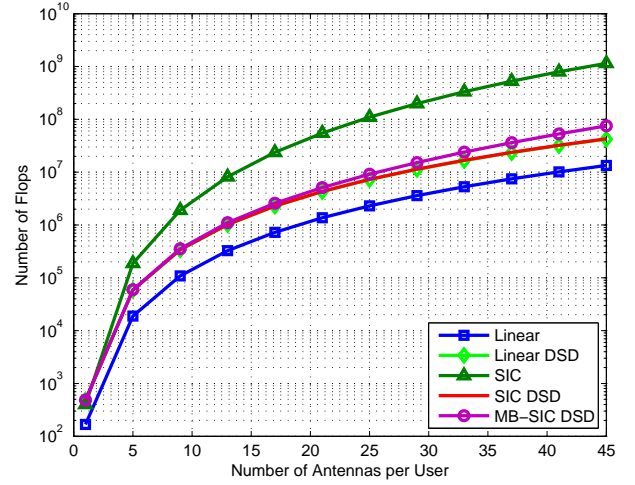}
\end{center}
\caption{Computational complexity versus number of antennas per user,  $K=10$ active user, $N=10$ classes of users, $N_t=KN_{t_{k,n}}$  transmit antennas and $N_r=2N_t$ receive antennas.}
\label{CC3}
\end{figure}

Fig.~\ref{CC1} shows the computational complexity versus the number of user classes for different detection algorithms. We consider $K=100$ active users, $N_{t_{k,n}}=2$ transmit antennas per user and $N_r=200$ receive antennas distributed around the cell. We consider an increasing number of classes of users, when $K$ is not divisible by the number of classes, the number of active users is set to a smaller value so as to allow the division in $N$ classes. We can see from the figure that the complexity of the SIC and the MB-SIC detectors with the DSD algorithm is lower than the SIC and the MB-SIC coupled detectors, respectively. Furthermore, for receivers with DSD the computational complexity is reduced as the number of user classes is increased. This fact represents an important advantage for receivers with DSD, due to the fact that it allows to use of more complex detectors for each user class according to its data requirements.

In~Fig.~\ref{CC2} and Fig.~\ref{CC3} we plot the required number of FLOPs versus the number of active users and versus the number of antennas per user, respectively. For Fig.~\ref{CC2} we consider $N=5$ classes of users, $N_{t_{k,n}}=2$ transmit antennas per user and $N_r=3KN_{t_{k,n}}$ receive antennas. The MB-SIC and SIC detectors with DSD has a lower computational cost than the coupled SIC detector. For this 5G context, with a high number of antennas, efficient coupled detectors are not feasible to be implemented, however, if a specific user class requires the benefits of complex detectors, the DSD algorithm reduces the cost so that more complex detectors could be applied as illustrated with MB-SIC in Fig.~\ref{CC2}. For the results in Fig.~\ref{CC3} we consider that we have $K=10$ active users and that we need to detect each user independently, i.e., $N=10$. The number of transmit antennas per user $N_{t_{k,n}}$ is increased. We also consider that the number of receive antennas distributed around the cell is given by $N_r=2KN_{t_{k,n}}$. The MB-SIC and SIC detectors with DSD have a significantly lower complexity when compared with the SIC detector where all users are coupled.

It is worth noting that the curves displayed in Fig.~\ref{CC1}, Fig.~\ref{CC2} and Fig.~\ref{CC3} will have a substantial decrease if the channel does not change over a time period due to quasi static channels. In this case, the equivalent channel matrices for each user class are stored for subsequent use.
It would increase the gap, in terms of the computational cost, for the detection schemes using the DSD algorithm.

\subsection{BER Performance}
In this subsection, the BER performance of the proposed DSD algorithm is evaluated using different detectors which includes linear, SIC and MB-SIC with linear MMSE receive filters.
The SIC detector of~\cite{ref2} uses a norm-based cancellation ordering, the MB-SIC of~\cite{ref7} employs a fixed number of branches, equal to the total number of transmit antennas per user class $N_{t_n}$ for the DSD schemes, and norm-based ordering in its first branch.
A massive MIMO system operating in heterogeneous networks with $K$ active users is considered. We also consider the DAS configuration where the $N_r=N_B+DQ$ receive antennas are distributed around the cell in $D$ radio heads with $Q$ antennas each and the remaining $N_B$ antennas are located at the BS. We consider QPSK modulation.
The SNR per transmitted information bit is defined as
\begin{equation}
\text{SNR}=10\log_{10} \frac{N_t\sigma_s^2}{M\sigma_n^2},
\end{equation}
where $\sigma_s^2$ is the common variance of the transmitted symbols, $\sigma_n^2$ is the noise variance at the receiver and  $M$ is the number of transmitted bits per symbol.
The numerical results correspond to an average of 3,000 simulations runs, with 500$N_t$ symbols transmitted per run. For the $N_B$ antennas at the BS, we employ $L_k=0.3$, $\tau=2$, the distance to the users is obtained from a uniform discrete random variable distributed between 0.4 and 0.7, the shadowing spread is $\sigma_k=1$ dB and the transmit and receive correlation coefficients are equal to $\rho_{rx}=0.4$. For the $R$ remote arrays of antennas, we use $L_{k,j}$ taken from a uniform random variable distributed between 0.3 and 0.5, the shadowing spread $\sigma_{k,j}=1$ dB and the receive correlation coefficients are equal to $\rho_{rx}=0.5$. When the number of transmit antennas at the users is $N_{t_{k,n}}>1$, the transmit correlation coefficient is equal to $\rho_{tx}=0.55$.

\begin{figure}[]
\begin{center}
\includegraphics[scale=0.62]{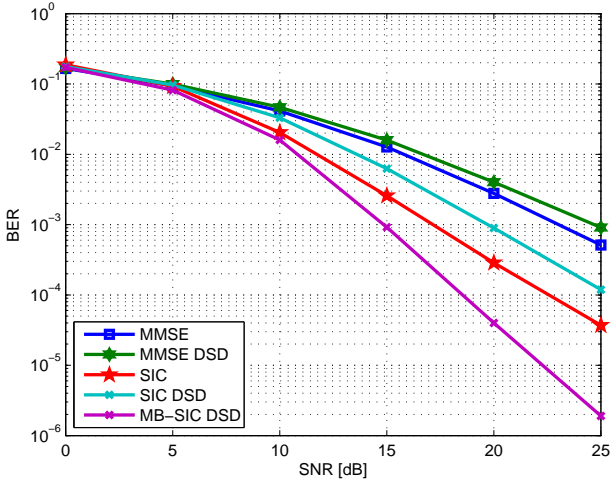}
\end{center}
\caption{BER versus SNR with  $K=12$ active users, $N=3$ classes of users, $N_{t_{k,n}}=3$  transmit antennas per user and $N_r=36$ receive antennas.}
\label{BER1}
\end{figure}

\begin{figure}[]
\begin{center}
\includegraphics[scale=0.62]{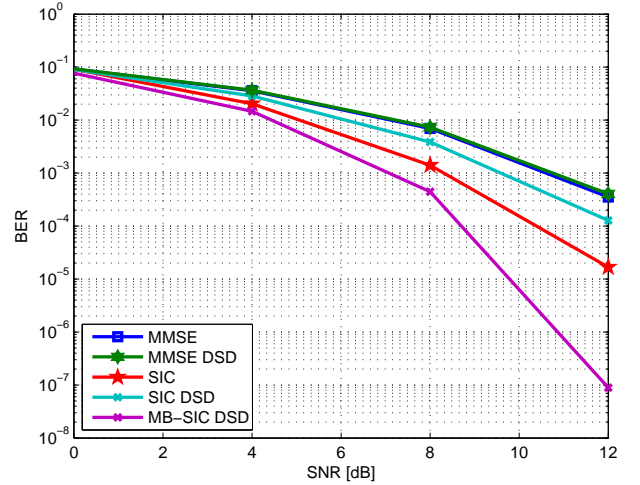}
\end{center}
\caption{BER versus SNR with $K=8$ active users, $N_{t_{k,n}}=8$  transmit antennas per user, $N=8$ classes of users and $N_r=128$ receive antennas.}
\label{BER2}
\end{figure}

For the experiment in~Fig.~\ref{BER1}, we consider $K=12$ user devices, where each user is equipped with $N_{t_{k,n}}=3$ transmit antennas and $N=3$ classes of users. The system employs perfect channel state information and QPSK modulation. For the DAS configuration, we consider $N_B=8$ receive antennas at the BS, $D=4$ remote radio heads and $Q=7$ receive antennas per remote radio head. We can see from the figure that the decoupled SIC detection presents a performance close to the coupled SIC detector with a difference around  2 dB in the high SNR region. In addition the decoupled SIC detector has a drastic reduction in the computational cost when compared with its coupled version.
The result also indicates a remarkable superiority in the performance for the MB-SIC receiver with the DSD scheme over the coupled SIC detector which also has a lower computational complexity than the coupled SIC detector.

\begin{figure}
\begin{center}
\includegraphics[scale=0.62]{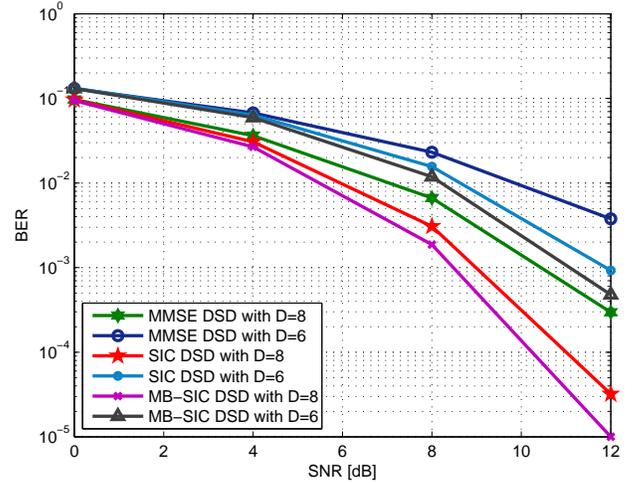}
\end{center}
\caption{BER versus SNR with  $K=64$ active users, $N_{t_{k,n}}=1$  transmit antennas per user, $N=4$ classes of users, $N_B=34$, $D$ remote antennas arrays with $Q=7$ receive antennas per array.}
\label{BER3}
\end{figure}

In the next example, we evaluate the performance of the proposed DSD
scheme considering $K=8$ active users, each user transmitting with
$N_{t_{k,n}}=8$ antennas. We also consider that we need to detect
each user independently of each other, i.e., $N=8$ classes of users
with $\mid C_n\mid=1$ users per class. For the DAS configuration, we
consider $N_B=64$ receive antennas at the BS, $D=8$ remote radio
heads and $Q=8$ receive antennas per remote radio head. The
Fig.~\ref{BER2} indicate that the performance of the SIC detector
with DSD is close to the SIC detector with a lower computational
complexity. Note that the results for MB-SIC with DSD show a very
good performance with a computational complexity much lower than the
SIC detector without DSD.

In Fig.~\ref{BER3} we evaluate the performance of the proposed DSD
scheme with MMSE, SIC and MB-SIC detection. We consider $K=64$
active users, $N=4$ classes of users, $\mid C_n \mid=16$ users per
class and $N_{t_{k,n}}=1$ transmit antennas per user. For the DAS
configuration, we consider $N_B=34$ receive antennas at the BS and
$Q=7$ receive antennas per remote radio head. To show the behavior
of the BER performance with different numbers of distributed
antennas we consider two configurations of remote radios heads,
$D=8$ and $D=6$. As expected, the results shows that when the number
of RRHs is increased, the BER performance is improved due to the low
propagation effects caused by the short distances between users and
some remote antenna arrays. We also can see from the figure that the
SIC and the MB-SIC detector with DSD offer an excellent BER
performance with a low computational cost.

\section{Conclusions}
\label{sec:conclusions}

In this paper a mathematical signal model and the DSD algorithm for
the uplink of massive MIMO systems operating in heterogeneous
cellular networks with different classes of users using CAS and  DAS
configurations have been presented. The proposed DSD allows one to
separate the received signals for each category of users efficiently
into independent parallel single user class signals at the receiver
side, applying a common channel inversion and QR decomposition and
assuming that the channel matrix was previously estimated. With the
proposed DSD scheme, it is possible to handle different classes of
users in heterogeneous networks and to use different modulation
and/or detection schemes for each class according to its data
service requirements. The main advantage of DSD is the reduction in
the computational cost of efficient detection schemes that, for its
high computational complexity, are not feasible for implementation
when the signals received from all active users are coupled.

\appendices
\section{}
\label{Appendix-1}
Let us consider that a linear detector according to~(\ref{eq20}) is applied to the equivalent received signal vector~(\ref{eq12}) to detect the transmitted symbol vector of the $n$-th user class as in~(\ref{24}).
If we define the matrix $\bar{\mathbf{A}}_n=\mathbf{W}_n\check{\mathbf{H}}_n$ and the vector $\bar{\mathbf{n}}_n=\mathbf{W}_n\mathbf{n}_n$ we can rewrite~(\ref{24}) as
\begin{equation}
\tilde{\mathbf{y}}_n=\bar{\mathbf{A}}_n\mathbf{s}_n+\bar{\mathbf{n}}_n,
\end{equation}
then, in analogy with the analysis in Section~IV, the sum rate for the $n$-th user class after the detection is given by
\begin{equation}
R_n=\sum_{i=1}^{N_{t_n}}\log_2(1+\bar{\lambda_i}),
\end{equation}
where the values $\bar{\lambda_i}$ are the eigenvalues of the matrix $\bar{\mathbf{B}}_n^\mathcal{H}\bar{\mathbf{B}}_n$ with $\bar{\mathbf{B}}_n=\mathbf{K}_{\bar{\mathbf{n}}_n}^{-1/2}\bar{\mathbf{A}}_n\mathbf{K}_{{\mathbf{s}}_n}^{1/2}$. Then
\begin{equation}
\bar{\mathbf{B}}_n^\mathcal{H}\bar{\mathbf{B}}_n=\mathbf{K}_{{\mathbf{s}}_n}^{1/2}\bar{\mathbf{A}}_n^\mathcal{H}\mathbf{K}_{\bar{\mathbf{n}}_n}^{-1}\bar{\mathbf{A}}_n\mathbf{K}_{{\mathbf{s}}_n}^{1/2},
\end{equation}
where $\mathbf{K}_{{\mathbf{s}}_n}=\sigma_s^2\mathbf{I}$ and $\mathbf{K}_{\bar{\mathbf{n}}_n}=\mathbf{W}_n\mathbf{K}_{{\mathbf{n}}_n} \mathbf{W}_n^\mathcal{H}$. Assuming that $\mathbf{K}_{\bar{\mathbf{n}}_n}$ has an inverse, we finally obtain
\begin{eqnarray}
\bar{\mathbf{B}}_n^\mathcal{H}\bar{\mathbf{B}}_n&=&\sigma_s^2\bar{\mathbf{A}}_n^\mathcal{H}(\mathbf{W}_n\mathbf{K}_{{\mathbf{n}}_n} \mathbf{W}_n^\mathcal{H})^{-1}\bar{\mathbf{A}}_n\nonumber \\
&=& \sigma_s^2 \check{\mathbf{H}}_n^\mathcal{H}\mathbf{W}_n^\mathcal{H} (\mathbf{W}_n\mathbf{K}_{{\mathbf{n}}_n} \mathbf{W}_n^\mathcal{H})^{-1}\mathbf{W}_n\check{\mathbf{H}}_n\nonumber \\
&=&\sigma_s^2 \check{\mathbf{H}}_n^\mathcal{H}\mathbf{K}_{{\mathbf{n}}_n} ^{-1}\check{\mathbf{H}}_n.\label{eqA2}
\end{eqnarray}
Note that~(\ref{eqA2}) and~(\ref{eq18}) will yield the same result which proves that the sum rate for the DSD algorithm is independent of the linear detection procedure.

\bibliographystyle{IEEEtran}
\bibliography{Referens}
\normalfont

\end{document}